\documentclass[conference]{IEEEtran}
\usepackage{amsfonts}

\IEEEoverridecommandlockouts

\usepackage{cite}
\usepackage{graphicx}
\usepackage[cmex10]{amsmath}
\usepackage{algorithmic}
\usepackage{algorithm}
\usepackage{array}
\usepackage{mdwmath}
\usepackage{mdwtab}
\usepackage[caption=false]{caption}
\usepackage[caption=false,font=footnotesize]{subfig}
\usepackage{fixltx2e}
\usepackage{stfloats}
\ifCLASSOPTIONcaptionsoff
 \usepackage[nomarkers]{endfloat}
 \let\MYoriglatexcaption\caption
 \renewcommand{\caption}[2][\relax]{\MYoriglatexcaption[#2]{#2}}
\fi
\usepackage{url}

\begin{document}
%
\title{On the Cram\'{e}r-Rao Lower Bound for Spatial Correlation Matrices of Doubly Selective Fading Channels for MIMO OFDM Systems}

\author{\IEEEauthorblockN{Xiaochuan~Zhao, Tao~Peng, Ming~Yang and Wenbo~Wang}%
\IEEEauthorblockA{Wireless Signal Processing and Network Lab \\
Key Laboratory of Universal Wireless Communication, Ministry of Education \\
Beijing University of Posts and Telecommunications, Beijing, China \\
Email: zhaoxiaochuan@gmail.com}%
\thanks{This work is sponsored in
part by the National Natural Science Foundation of China under grant
No.60572120 and 60602058, and in part by the national high
technology researching and developing program of China (National 863
Program) under grant No.2006AA01Z257 and by the National Basic
Research Program of China (National 973 Program) under grant
No.2007CB310602.}}

\maketitle

\begin{abstract}
In this paper, the Cram\'er-Rao lower bound (CRLB) for spatial
correlation matrices is derived based on a rigorous model of the
doubly selective fading channel for multiple-input multiple-output
(MIMO) orthogonal frequency division multiplexing (OFDM) systems.
Adopting an orthogonal pilot pattern for multiple transmitting
antennas and assuming independent samples along the time, the sample
auto-correlation matrix of the channel response is complex Wishart
distributed. Then, the maximum likelihood estimator (MLE) and the
analytic expression of CRLB are derived by assuming that temporal
and frequency correlations are known. Furthermore, lower bounds of
total mean squared error (TMSE) and average mean squared error
(AvgMSE) are deduced from CRLB for asymptotically infinite and
finite signal-to-noise ratios (SNR's), respectively. According to
the lower bound of AvgMSE, the amount of samples and the order of
frequency selectivity show dominant impact on the accuracy of
estimation. Besides, the number of pilot tones, SNR and normalized
maximum Doppler spread together influence the effective order of
frequency selectivity. Numerical simulations demonstrate the
analytic results.
\end{abstract}


\renewcommand{\thefootnote}{\fnsymbol{footnote}}

\section{Introduction}
\label{sec:intro} Due to the virtues of orthogonal frequency
division multiplexing (OFDM) for converting frequency selective
fading channels into flat fading ones and of multiple-input
multiple-output (MIMO) techniques for exploiting spatial diversity
gain and/or enhancing the system capacity, many current systems
combine them two together to achieve a better quality as well as a
higher throughput \cite{Stub04}.

For MIMO systems, the spatial correlation matrices play very
important roles and are widely utilized, for example, to facilitate
the transmitting precoding \cite{Goro00}\cite{Samp01}, MMSE receiver
\cite{Nord06} and multi-user strategies \cite{Jors03}. However,
since the true spatial correlation matrices are unknown in real
applications, the sample correlation matrices have to be used in
stead and are usually obtained through the channel estimation.

In this paper, we study the Cram\'er-Rao lower bound (CRLB) for the
sample spatial correlation matrices of doubly selective fading
channels for MIMO OFDM systems. Based on the a rigorous doubly
fading channel model and assuming invariant pilot sequence along the
time, the maximum likelihood estimator (MLE) and the CRLB are
derived. Then, the analytic expressions of lower bounds of the total
mean squared error (TMSE) and average mean squared error (AvgMSE)
are obtained for asymptotically infinite and finite signal-to-noise
ratios (SNR's), respectively. Based on the lower bound of AvgMSE,
several factors influencing the accuracy of estimation, including
the amount of samples, the order of frequency selectivity, the
normalized maximum Doppler spread, the number of pilot tones per
antenna and SNR, are further analyzed.

This paper is organized as follows. In Section \ref{sec:model}, the
MIMO OFDM system and channel model are introduced. Then, in Section
\ref{sec:CRLB}, CRLB of the sample spatial correlation matrix is
derived and further lower bounded to uncover the essential factors.
Numerical results appear in Section \ref{sec:numresults}. Finally,
Section \ref{sec:conclusion} concludes the paper.

\emph{Notation}: Lowercase and uppercase boldface letters denote
column vectors and matrices, respectively. $(\cdot)^*$, $(\cdot)^T$,
$(\cdot)^H$, $(\cdot)^{\dagger}$, and $||\cdot||_F$ denote
conjugate, transposition, conjugate transposition, Moore-Penrose
pseudo-inverse and Frobenius norm, respectively. $\otimes$ denotes
the Kronecker product. $E(\cdot)$ represents expectation.
$[{\bf{A}}]_{i,j}$ and $[{\bf{a}}]_{i}$ denotes the $(i$,$j)$-th
element of ${\bf{A}}$ and the $i$-th element of ${\bf{a}}$,
respectively. ${\text{diag}}({\bf{a}})$ is a diagonal matrix by
placing ${\bf{a}}$ on the diagonal.

\setlength{\arraycolsep}{0.2em}

\section{System Model}
\label{sec:model} Consider an MIMO OFDM system with a bandwidth of
$BW=1/T$ Hz ($T$ is the sampling period). $N$ denotes the total
number of tones, and a cyclic prefix (CP) of length $L_{cp}$ is
inserted before each symbol to eliminate inter-block interference.
Thus the whole symbol duration is $T_s=(N+L_{cp})T$. Then, $n_T$
transmitting antennas at the base station (BS) and $n_R$ receiving
antennas at the mobile station (MS) are assumed, respectively.

Between the $i$-th transmitting antenna and $j$-th receiving
antenna, the complex baseband model of a linear time-variant mobile
channel with $L^{(j,i)}$ paths can be described by \cite{Stee92}
\begin{equation}
\label{eqn:channel}
{h^{(j,i)}(t,\tau)=\sum\limits_{l=1}^{L^{(j,i)}}{h_l^{(j,i)}(t)\delta\left({\tau-\tau_l^{(j,i)}T}\right)}}
\end{equation}
where $(\tau_l)^{(j,i)}\in{\mathcal{R}}$ is the normalized
non-sample-spaced delay of the $l$-th path, and $h_l^{(j,i)}(t)$ is
the corresponding complex amplitude with the power
$({{\sigma}_l^2})^{(j,i)}$.

The following conditions are assumed to characterize the correlation
property of the channel.
\begin{enumerate}
  \item \label{item:space} Space Correlation: the stochastic MIMO radio channel model \cite{Kerm02} is
  adopted;
  \item \label{item:freq} Frequency Correlation: the wide-sense stationary uncorrelated scattering
(WSSUS) \cite{Stee92} is assumed;
  \item \label{item:time} Time Correlation: the uniform scattering environment
introduced by Clarke \cite{Clar68} is assumed.
  \item \label{item:indep} Scattering Function Separability: the channel has degeneracy in all three
  dimensions \cite{Paul03}.
\end{enumerate}

Therefore, the complete spatial correlation matrix of the MIMO radio
channel is given by \cite{Kerm02}
\begin{equation}
\label{eqn:corspacedef}
{\bf{\Xi}}_s={\bf{\Xi}}_{s,T}\otimes{\bf{\Xi}}_{s,R}
\end{equation}
where ${\bf{\Xi}}_{s,T}\in{\mathcal{C}}^{n_T\times{n_T}}$ and
${\bf{\Xi}}_{s,R}\in{\mathcal{C}}^{n_R\times{n_R}}$ are the
symmetrical complex correlation matrices of antenna arrays of BS and
MS, respectively. Then, the normalized time correlation function of
any path is identical, i.e., \cite{Clar68}
\begin{equation}
\label{eqn:rtdef}
{r_{t}({\Delta}t)=J_0\left(2{\pi}{f_d}{\Delta}t\right)}
\end{equation}
where $f_d$ is the maximum Doppler spread, and $J_0(\cdot)$ is the
zeroth order Bessel function of the first kind. In addition,
$L^{(j,i)}=L$, $(\tau_l)^{(j,i)}=\tau_l$ and
$({{\sigma}_l^2})^{(j,i)}={\sigma}_l^2$, hence, the frequency
correlation matrix is
\begin{equation}
\label{eqn:Rfdef}
{\bf{R}}_{f}={\bf{F}}_{\tau}{\bf{D}}{\bf{F}}_{\tau}^H
\end{equation}
where ${\bf{D}}={\text{diag}}({\sigma}_l^2)$, $l=1,\ldots,L$ and
${\bf{F}}_{\tau}\in{\mathcal{C}}^{N\times{L}}$ is the unbalanced
Fourier transform matrix, defined as
$[{\bf{F}}_{\tau}]_{k,l}=e^{-j2{\pi}k{\tau}_l/N}$. Moreover, the
power of the channel between each pair of transmitting and receiving
antennas is normalized, i.e., ${\text{tr}}({\bf{D}})=1$.

Assuming a sufficient CP, i.e., $L_{cp} \geq L$, the discrete signal
model in the frequency domain is written as
\begin{equation}
\label{eqn:YmMatrixdef}
{\bf{y}}_f^{(j)}(n)=\sum\limits_{i=1}^{n_T}{\bf{H}}_f^{(j,i)}(n){\bf{x}}_f^{(i)}(n)+{\bf{n}}_f^{(j)}(n)
\end{equation}
where
${\bf{x}}_f^{(i)}(n),{\bf{y}}_f^{(j)}(n),{\bf{n}}_f^{(j)}(n)\in{\mathcal{C}}^{N\times1}$
are the $n$-th transmitted and received signal and additive white
Gaussian noise (AWGN) vectors, respectively, and
${\bf{H}}_f^{(j,i)}(n)\in{\mathcal{C}}^{N\times{N}}$ is the channel
transfer matrix between the $i$-th transmitting and $j$-th receiving
antennas with the $(k+\nu,k)$-th element as
\begin{equation}
\label{eqn:Hdef}
\left[{\bf{H}}_f^{(j,i)}(n)\right]_{k+\upsilon,k}=\frac{1}{N}\sum\limits_{m=0}^{N-1}\sum\limits_{l=1}^{L}h_l^{(j,i)}(n,m)e^{-j2{\pi}({\upsilon}m+k{\tau_l})/N}
\end{equation}
where $h_l^{(j,i)}(n,m)=h_l^{(j,i)}(nT_s+(L_{cp}+m)T)$ is the
sampled complex amplitude of the $l$-th path. $k$ and $\upsilon$
denote frequency and Doppler spread, respectively.

\section{CRLB of Spatial Correlation Matrices}
\label{sec:CRLB}  Usually the correlation matrices of the channel
response are obtained through the least squared (LS) channel
estimation on pilot tones, that is, only pilot symbols are extracted
and used to perform LS channel estimation. Therefore,
\begin{equation}
\label{eqn:lsdef}
{\bf{h}}_{f,ls}^{(j,i)}(n)=({\bf{X}}_{f}^{(i)}(n))^{-1}{\bf{y}}_{f}^{(j)}(n)
\end{equation}
where ${\bf{X}}_{f}^{(i)}(n)={\text{diag}}({\bf{x}}_f^{(i)}(n))$ is
a diagonal matrix consisting of pilot symbols. To alleviate the
interference between multiple transmitting antennas, pilot symbols,
i.e., ${\bf{x}}_f^{(i)}(n)$, $i=1,\ldots,n_T$, are designed to be
orthogonal, therefore
\begin{equation}
\label{eqn:orthpilot}
({\bf{x}}_f^{(i_1)}(n))^H{\bf{x}}_f^{(i_2)}(n)={\delta}(i_1-i_2)|{\bf{x}}_f^{(i_1)}(n)|_2^2
\end{equation}
In this paper, we adopt the frequency division pilot pattern of
which the pilot tones allocated to a certain transmitting antenna is
exclusive of the others. To be more specifically, it is assumed that
the pilot tone set of the $i$-th transmitting antenna is
${\mathcal{I}}_{p}^{(i)}=\{i+k{\times}{\theta};k=0,\ldots,P-1\}$,
where $P$ is the number of pilot tones of each antenna satisfying
$L{\leq}P$, $\theta{\geq}n_T$ and $P\theta\leq{N}$. Thus, the $i$-th
transmitting antenna transmits non-zero pilots on
${\mathcal{I}}_{p}^{(i)}$ while nulls on the rest. Besides, we
assume that the non-zero elements of ${\bf{x}}_f^{(i)}(n)$
constitute a fixed vector ${\bf{x}}_p\in{\mathcal{C}}^{P\times1}$,
which is independent of $n$ and $i$ and of the normalized power so
that ${\bf{X}}_{p}{\bf{X}}_{p}^H={\bf{I}}_P$, where
${\bf{X}}_{p}={\text{diag}}({\bf{x}}_{p})$. Then, by discarding the
zero elements, (\ref{eqn:lsdef}) is further rewritten into
\begin{equation}
\label{eqn:lsdefnew}
{\bf{h}}_{p,ls}^{(j,i)}(n)={\bf{X}}_{p}^{-1}{\bf{H}}_p^{(j,i)}(n){\bf{x}}_p+{\bf{X}}_{p}^{-1}{\bf{n}}_p^{(j,i)}(n)
\end{equation}
where the subscript $p$ denotes elements on pilot tones.

According to the orthogonal pilot pattern, the instantaneous channel
impulse response (CIR) vector between the $i$-th transmitting
antenna and the $j$-th receiving antenna corresponding to the
$(i+k{\times}{\theta})$-th sample of the $n$-th OFDM symbol can be
denoted as
${\bf{h}}_t^{(j,i)}(n,i+k{\times}{\theta})=[h_1^{(j,i)}(n,i+k{\times}{\theta}),\ldots,h_L^{(j,i)}(n,i+k{\times}{\theta})]^T$,
$k=0,\ldots,P-1$. Then, the $(j,i)$-th CIR matrix for the $n$-th
OFDM symbol is formed as
${\bf{H}}_t^{(j,i)}(n)=[{\bf{h}}_t^{(j,i)}(n,i),\ldots,{\bf{h}}_t^{(j,i)}(n,i+(P-1)\times\theta)]$.
According to the assumptions of WSSUS and uniform scattering,
${\bf{H}}_t^{(j,i)}(n)$ is complex normal, i.e.,
\begin{equation}
\label{eqn:Ht}
{\bf{H}}_t^{(j,i)}(n)\;{\sim}\;{\mathcal{CN}}_{L\times{P}}(0,{\bf{\Omega}}\otimes{\bf{D}})
\end{equation}
where ${\bf{\Omega}}\in{\mathcal{C}}^{P\times{P}}$ is a Toeplitz
matrix, defined as
\begin{equation}
\label{eqn:Omegadef}
[{\bf{\Omega}}]_{k_1,k_2}=J_0\left(2{\pi}{f_d}(k_1-k_2){\theta}T\right)
\end{equation}

Then according to (\ref{eqn:Hdef}) and the pilot pattern, the
$(j,i)$-th channel transfer matrix is
${\bf{H}}_p^{(j,i)}(n)={\bf{F}}_{\tau}^{(i)}{\bf{H}}_t^{(i)}(n)$,
where ${\bf{F}}_{\tau}^{(i)}$ is the submatrix of ${\bf{F}}_{\tau}$
by drawing rows from ${\mathcal{I}}_{p}^{(i)}$. Thus
\begin{equation}
\label{pdf:Hf}
{\bf{H}}_p^{(j,i)}(n)\;{\sim}\;{\mathcal{CN}}_{P\times{P}}(0,{\bf{\Omega}}\otimes[{\bf{F}}_{\tau}^{(i)}{\bf{D}}({\bf{F}}_{\tau}^{(i)})^H])
\end{equation}
Assuming CIR is independent of the thermal noise, with
(\ref{eqn:lsdefnew}) and (\ref{pdf:Hf}), we have
\begin{equation}
\label{pdf:hls}
{\bf{h}}_{p,ls}^{(j,i)}(n)\;{\sim}\;{\mathcal{CN}}_{P}(0,{\bf{\Upsilon}}^{(j,i)})
\end{equation}
where the covariance matrix ${\bf{\Upsilon}}^{(j,i)}$ is defined as
\begin{equation}
\label{eqn:Sigmadef}
{\bf{\Upsilon}}^{(j,i)}=({\bf{x}}_{p}^H{\bf{\Omega}}{\bf{x}}_{p})({\bf{X}}_{p}^{H}{\bf{F}}_{\tau}^{(i)}{\bf{D}}({\bf{F}}_{\tau}^{(i)})^H{\bf{X}}_{p})+{\sigma}_n^2{\bf{I}}_{P}
\end{equation}
Apparently, ${\bf{\Upsilon}}^{(j,i)}$ is irrelevant of the indexes
of the receiving antennas, which follows the assumption
\ref{item:indep}). According to the orthogonal pilot pattern,
${\bf{F}}_{\tau}^{(i)}={\bf{F}}_{\tau}^{(0)}{\bf{\Phi}}^{i}$, where
${\bf{F}}_{\tau}^{(0)}\in{\mathcal{C}}^{P\times{L}}$ is a submatrix
of ${\bf{F}}_{\tau}$ by drawing rows from
${\mathcal{I}}_{p}^{(0)}=\{k\times\theta;k=0,\ldots,P-1\}$ and
${\bf{\Phi}}$ is a diagonal phase-twisted matrix with
$[{\bf{\Phi}}]_{l,l}=e^{-j2{\pi}{\tau_l}/N}$. Then, the frequency
auto-correlation matrix of ${\mathcal{I}}_{p}^{(i)}$ is
\begin{equation}
\label{eqn:subRfdef}
{\bf{R}}_{p}^{(i)}={\bf{F}}_{\tau}^{(i)}{\bf{D}}({\bf{F}}_{\tau}^{(i)})^H={\bf{F}}_{\tau}^{(0)}{\bf{\Phi}}^{i}{\bf{D}}({\bf{F}}_{\tau}^{(0)}{\bf{\Phi}}^{i})^H={\bf{R}}_{p}
\end{equation}
where
${\bf{R}}_{p}={\bf{F}}_{\tau}^{(0)}{\bf{D}}({\bf{F}}_{\tau}^{(0)})^H$.
Furthermore, to simplify the following derivation, we assume that
CFR varies slightly within $n_T$ contiguous tones, so that the
frequency cross-correlation matrix between the pilot sets
${\mathcal{I}}_{p}^{(i_1)}$ and ${\mathcal{I}}_{p}^{(i_2)}$ is
\begin{equation}
\label{eqn:crossRp}
{\bf{R}}_{p}^{(i_1,i_2)}={\bf{F}}_{\tau}^{(i_1)}{\bf{D}}({\bf{F}}_{\tau}^{(i_2)})^H={\bf{F}}_{\tau}^{(0)}{\bf{\Phi}}^{i_1-i_2}{\bf{D}}({\bf{F}}_{\tau}^{(0)})^H\approx{\bf{R}}_{p}
\end{equation}
Then, the complete LS-estimated channel transfer matrix is
constructed as
\begin{equation}
\label{eqn:completeHdef} {\mathbb{H}}_{p,ls}(n)=\left(
                      \begin{array}{ccc}
                        {\bf{h}}_{p,ls}^{(1,1)}(n) & \cdots & {\bf{h}}_{p,ls}^{(1,n_T)}(n) \\
                        \vdots & \ddots & \vdots \\
                        {\bf{h}}_{p,ls}^{(n_R,1)}(n) & \cdots & {\bf{h}}_{p,ls}^{(n_R,n_T)}(n) \\
                      \end{array}
                    \right)_{Pn_R\times{n_T}}
\end{equation}
and it is complex normal, i.e.,
\begin{equation}
\label{pdf:completeH}
{\mathbb{H}}_{p,ls}(n)\;{\sim}\;{\mathcal{CN}}_{Pn_R\times{n_T}}(0,{\bf{\Sigma}})
\end{equation}
where ${\bf{\Sigma}}\in{\mathcal{C}}^{Pn_Rn_T\times{Pn_Rn_T}}$.
According to the assumptions \ref{item:space})-\ref{item:indep}),
its $((i_1-1)n_R+j_1,(i_2-1)n_R+j_2)$-th submatrix is
(\ref{eqn:subSigma}), shown at bottom of the next page.
\begin{figure*}[!b]
\hrule \vspace{2pt} \normalsize {\begin{equation}
\label{eqn:subSigma}
\{{\bf{\Sigma}}\}_{(i_1-1)n_R+j_1,(i_2-1)n_R+j_2}^{(P\times{P})}=[{\bf{\Xi}}_s]_{(i_1-1)n_R+j_1,(i_2-1)n_R+j_2}\times(\omega{\bf{X}}_{p}^{H}{\bf{R}}_{p}^{(i_1,i_2)}{\bf{X}}_{p})+{\sigma}_n^2{\delta}(i_1-i_2){\delta}(j_1-j_2){\bf{I}}_{P}
\end{equation}}
\end{figure*}
Then, with (\ref{eqn:subRfdef}) and (\ref{eqn:crossRp}),
${\bf{\Sigma}}$ is expressed as
\begin{equation}
\label{eqn:completeSigma}
{\bf{\Sigma}}={\bf{\Xi}}_s\otimes(\omega{\bf{A}})+{\sigma}_n^2{\bf{I}}_{n_Tn_RP}
\end{equation}
where $\omega={\bf{x}}_{p}^H{\bf{\Omega}}{\bf{x}}_{p}$ and
${\bf{A}}={\bf{X}}_{p}^{H}{\bf{R}}_{p}{\bf{X}}_{p}$.

When $N_t$ complete LS estimated channel transfer matrices are
available, the sample auto-correlation matrix is formed as
\begin{equation}
\label{eqn:estRdef}
{\bf{\hat{\Sigma}}}=\frac{1}{N_{t}}\sum\limits_{n=1}^{N_{t}}{\text{vec}}({\mathbb{H}}_{p,ls}(n)){\text{vec}}({\mathbb{H}}_{p,ls}(n))^H
\end{equation}
where $N_t$ is the number of samples. To derive the probability
density function (PDF) of the sample auto-correlation matrix, we
assume that samples are independent of each other, which may be a
strict constraint. However, when the spacing between two contiguous
pilot symbols is sufficiently large, the correlation between them is
rather low, which alleviates the effect of model mismatch. Then,
${\bf{\hat{\Sigma}}}$ has the complex central Wishart distribution
with $N_t$ degrees of freedom and covariance matrix
${\bf{\Sigma}}'={\bf{\Sigma}}/{N_t}$ \cite{Ratn03}, denoted as
\begin{equation}
\label{pdf:Rls}
{\bf{\hat{\Sigma}}}\;{\sim}\;{\mathcal{CW}}_{n_Tn_RP}(N_t,{\bf{\Sigma}}')
\end{equation}
and its PDF is
\begin{equation}
\label{pdf:Rlsfunc}
f({\bf{\hat{\Sigma}}})=\frac{{\text{etr}}(-{\bf{\Sigma}}'^{-1}{\bf{\hat{\Sigma}}})[\det({\bf{\hat{\Sigma}}})]^{N_t-n_Tn_RP}}{{C\Gamma}_{n_Tn_RP}(N_t)[\det({\bf{\Sigma}}')]^{N_t}}
\end{equation}
where ${\text{etr}}(\cdot)=\exp({\text{tr}}(\cdot))$ and
${C\Gamma}_{n_Tn_RP}(N_t)$ is the complex multivariate gamma
function, defined as
\begin{equation}
\label{eqn:complexGamma}
{C\Gamma}_{n_Tn_RP}(N_t)={\pi}^{n_Tn_RP(n_Tn_RP-1)/2}\prod\limits_{k=1}^{n_Tn_RP}{\Gamma}(N_t-k+1)\nonumber
\end{equation}
Then, according to (\ref{pdf:Rlsfunc}), the likelihood function with
respect to the parameter matrix ${\bf{\Xi}}_s$ can be written as
\begin{eqnarray}
{\text{L}}({\bf{\Xi}}_s)&=&{\text{tr}}(-{\bf{\Sigma}}'^{-1}{\bf{\hat{\Sigma}}})+(N_t-n_Tn_RP)\ln(\det({\bf{\hat{\Sigma}}}))\nonumber\\
\label{likelihood:Rlsfunc}
{}&{}&-\ln({C\Gamma}_{n_Tn_RP}(N_t))-N_t\ln(\det({\bf{\Sigma}}'))
\end{eqnarray}
Therefore, the score function \cite{Mard79} is
\begin{equation}
\label{score:Rlsfunc}
{\text{score}}({\text{vec}}({\bf{\Xi}}_s))=\frac{\partial{{\text{L}}({\bf{\Xi}}_s)}}{\partial{{\text{vec}}({\bf{\Xi}}_s)}}=\frac{\partial{{\text{vec}}({\bf{\Sigma}}')^T}}{\partial{{\text{vec}}({\bf{\Xi}}_s)}}\frac{\partial{{\text{L}}({\bf{\Xi}}_s)}}{\partial{{\text{vec}}({\bf{\Sigma}}')}}
\end{equation}
where the first multiplicative term on the right-hand side,
according to (\ref{eqn:completeSigma}), is
\begin{equation}
\label{eqn:scorefirstterm}
\frac{\partial{{\text{vec}}({\bf{\Sigma}}')^T}}{\partial{{\text{vec}}({\bf{\Xi}}_s)}}=\frac{\omega}{N_t}\frac{\partial{{\text{vec}}({\bf{\Xi}}_s\otimes{\bf{A}})^T}}{\partial{{\text{vec}}({\bf{\Xi}}_s)}}
\end{equation}
and the second term is
\begin{equation}
\label{eqn:scoresecondterm}
\frac{\partial{{\text{L}}({\bf{\Xi}}_s)}}{\partial{{\text{vec}}({\bf{\Sigma}}')}}={\text{vec}}[({\bf{\Sigma}}'^{-1}{\bf{\hat{\Sigma}}}{\bf{\Sigma}}'^{-1}-N_t{\bf{\Sigma}}'^{-1})^T]
\end{equation}
Since
\begin{equation}
\label{eqn:vecidentity}
{\text{vec}}({\bf{\Xi}}_s\otimes{\bf{A}})={\bf{K}}_{\otimes}[{\text{vec}}({\bf{\Xi}}_s)\otimes{\text{vec}}({\bf{A}})]
\end{equation}
where ${\bf{K}}_{\otimes}$ is
\begin{equation}
\label{eqn:Kdef}
{\bf{K}}_{\otimes}={\bf{I}}_{n_Rn_T}\otimes{\bf{K}}_{(n_Rn_T)P}^T\otimes{\bf{I}}_{P}
\end{equation}
where
${\bf{K}}_{(n_Rn_T)P}\in{\mathcal{R}}^{(n_Rn_T)P{\times}(n_Rn_T)P}$
is a transpose matrix satisfying
${\text{vec}}({\bf{B}}^T)={\bf{K}}_{(n_Rn_T)P}{\text{vec}}({\bf{B}})$,
where ${\bf{B}}\in{\mathcal{C}}^{n_Rn_T{\times}P}$. Hence,
(\ref{eqn:scorefirstterm}) is rewritten into
\begin{equation}
\label{eqn:scorefirsttermnew}
\frac{\partial{{\text{vec}}({\bf{\Sigma}}')^T}}{\partial{{\text{vec}}({\bf{\Xi}}_s)}}=\frac{\omega}{N_t}[{\bf{I}}_{(n_Rn_T)^2}\otimes{\text{vec}}({\bf{A}})^T]{\bf{K}}_{\otimes}^T
\end{equation}

By letting the score function equal zero, we know that the maximum
likelihood estimator (MLE) of ${\bf{\Sigma}}$ is
${\bf{\hat{\Sigma}}}$, so, with (\ref{eqn:subSigma}), its
$(k_1,k_2)$-th submatrix, denoted as
$\{{\bf{\Sigma}}\}_{k_1,k_2}^{(P\times{P})}\in{\mathcal{C}}^{P\times{P}}$,
can be used to estimate $[{\bf{\Xi}}_s]_{k_1,k_2}$ by
\begin{equation}
\label{eqn:MLEfirst}
[{\bf{\Xi}}_s]_{k_1,k_2}\times(\omega{\bf{A}}_{k_1,k_2})+{\sigma_n^2}{\delta}(k_1-k_2){\bf{I}}_P=\{{\bf{\Sigma}}\}_{k_1,k_2}^{(P\times{P})}
\end{equation}
where $k_1=(i_1-1)n_R+j_1$, $k_2=(i_2-1)n_R+j_2$, and
\begin{equation}
\label{eqn:Ak1k2def}
{\bf{A}}_{k_1,k_2}={\bf{X}}_{p}^{H}{\bf{R}}_{p}^{(i_1,i_2)}{\bf{X}}_{p}
\end{equation}
Note that ${\bf{A}}_{k_1,k_2}$ and ${\sigma_n^2}$ are assumed to be
known for (\ref{eqn:MLEfirst}), that is, the frequency
auto/cross-correlation matrices, the pilot sequence and the noise
power are available. In order to solve $[{\bf{\Xi}}_s]_{k_1,k_2}$
from (\ref{eqn:MLEfirst}), the singular matrices of
${\bf{A}}_{k_1,k_2}$, denoted as ${\bf{U}}_{k_1,k_2}$ and
${\bf{V}}_{k_1,k_2}$, is used to transform (\ref{eqn:MLEfirst}) into
\begin{equation}
\label{eqn:MLEsecond}
\omega[{\bf{\Xi}}_s]_{k_1,k_2}{\bf{\Lambda}}_{k_1,k_2}+{\sigma_n^2}{\delta}(k_1-k_2){\bf{I}}_P={\bf{U}}_{k_1,k_2}^H\{{\bf{\Sigma}}\}_{k_1,k_2}^{(P\times{P})}{\bf{V}}_{k_1,k_2}\nonumber
\end{equation}
where ${\bf{\Lambda}}_{k_1,k_2}$ is a diagonal matrix with singular
values of ${\bf{A}}_{k_1,k_2}$ on the diagonal. Then, since
${\text{rank}}({\bf{A}}_{k_1,k_2})={\text{rank}}({\bf{R}}_{p})=L$,
the MLE of $[{\bf{\Xi}}_s]_{k_1,k_2}$ is
\begin{eqnarray}
{}&{}&\!\!\!\!\!\!\!\!{\text{MLE}}([{\bf{\Xi}}_s]_{k_1,k_2})=\nonumber\\
\label{eqn:MLEdef}
{}&{}&\!\!\!\!\!\sum\limits_{l=1}^{L}c_l\frac{[{\bf{U}}_{k_1,k_2}^H\{{\bf{\Sigma}}\}_{k_1,k_2}^{(P\times{P})}{\bf{V}}_{k_1,k_2}]_{l,l}-{\sigma_n^2}{\delta}(k_1-k_2)}{[{\bf{\Lambda}}_{k_1,k_2}]_{l,l}}
\end{eqnarray}
where $c_l$'s are normalized non-negative weight coefficients, i.e.,
$c_l\ge0$ and $\sum\nolimits_{l=1}^{L}c_l=1$.

Further, according to the score function, the Fisher Information
matrix with respect to ${\bf{\Xi}}_s$ \cite{Mard79} is
\begin{equation}
\label{eqn:fisherinfordef}
{\bf{J}}({\bf{\Xi}}_s)=E\left[\left(\frac{\partial{{\text{L}}({\bf{\Xi}}_s)}}{\partial{{\text{vec}}({\bf{\Xi}}_s)}}\right)\left(\frac{\partial{{\text{L}}({\bf{\Xi}}_s)}}{\partial{{\text{vec}}({\bf{\Xi}}_s)}}\right)^H\right]
\end{equation}
with
(\ref{score:Rlsfunc})(\ref{eqn:scorefirsttermnew})(\ref{eqn:scoresecondterm}),
(\ref{eqn:fisherinfordef}) is rewritten into
(\ref{eqn:fisherinfordefnew}), shown at the bottom of the next page,
where
${\bf{B}}={\bf{\Sigma}}'^{-1}{\bf{\hat{\Sigma}}}{\bf{\Sigma}}'^{-1}$.
\begin{figure*}[!b]
\vspace{2pt} \hrule \normalsize {\begin{equation}
\label{eqn:fisherinfordefnew}
{\bf{J}}({\bf{\Xi}}_s)=\frac{\omega^2}{N_t^2}[{\bf{I}}_{(n_Rn_T)^2}\otimes{\text{vec}}({\bf{A}})^T]{\bf{K}}_{\otimes}^TE\{{\text{vec}}[({\bf{B}}-N_t{\bf{\Sigma}}'^{-1})^T]{\text{vec}}[({\bf{B}}-N_t{\bf{\Sigma}}'^{-1})^T]^H\}{\bf{K}}_{\otimes}[{\bf{I}}_{(n_Rn_T)^2}\otimes{\text{vec}}({\bf{A}})^*]
\end{equation}}
\end{figure*}
Notice that
\begin{equation}
{\bf{B}}\;{\sim}\;{\mathcal{CW}}_{N}(N_t,{\bf{\Sigma}}'^{-1})\nonumber
\end{equation}
therefore
\begin{equation}
\label{eqn:secondterm}
\!\!E\{{\text{vec}}[({\bf{B}}-N_t{\bf{\Sigma}}'^{-1})^T]{\text{vec}}[({\bf{B}}-N_t{\bf{\Sigma}}'^{-1})^T]^H\}={\text{Var}}[{\text{vec}}({\bf{B}}^T)]
\end{equation}
According to \cite{Maiw00}, (\ref{eqn:secondterm}) is
\begin{equation}
\label{eqn:vardef}
{\text{Var}}[{\text{vec}}({\bf{B}}^T)]=N_t({\bf{\Sigma}}'^{-H}\otimes{\bf{\Sigma}}'^{-T})
\end{equation}
Then, with (\ref{eqn:vardef}), ${\bf{J}}({\bf{\Xi}}_s)$ is
\begin{eqnarray}
{\bf{J}}({\bf{\Xi}}_s)&=&\frac{\omega^2}{N_t}[{\bf{I}}_{(n_Rn_T)^2}\otimes{\text{vec}}({\bf{A}}_{k_1,k_2})^T]{\bf{K}}_{\otimes}^T({\bf{\Sigma}}'^{-H}\otimes{\bf{\Sigma}}'^{-T})\nonumber\\
\label{eqn:Fisherinformmatrix}
{}&{}&\times{\bf{K}}_{\otimes}[{\bf{I}}_{(n_Rn_T)^2}\otimes{\text{vec}}({\bf{A}}_{k_1,k_2})^*]
\end{eqnarray}
Therefore, the CRLB of ${\bf{\Xi}}_s$ is \cite{Mard79}
\begin{equation}
\label{eqn:CRLBdef}
{\text{CRLB}}({\bf{\Xi}}_s)={\bf{J}}^{-1}({\bf{\Xi}}_s)
\end{equation}

Now we consider the case that the SNR is asymptotically infinite,
or, equivalently, the power of noise is zero. According to
(\ref{eqn:completeSigma}), then, ${\bf{\Sigma}}'$ is reduced to
\begin{equation}
\label{eqn:Sigmaprimeinfinitesnr}
{\bf{\Sigma}}'=\frac{\omega}{N_t}{\bf{\Xi}}_s\otimes{\bf{A}}
\end{equation}
Since ${\bf{A}}$ is rank deficient if $L<P$, ${\bf{\Sigma}}'^{-1}$
should be replaced by ${\bf{\Sigma}}'^{\dagger}$. Then
\begin{equation}
\label{eqn:highsnrJ}
{\bf{K}}_{\otimes}^{T}({\bf{\Sigma}}'^{{\dagger}H}\otimes{\bf{\Sigma}}'^{{\dagger}T}){\bf{K}}_{\otimes}=\frac{N_t^2}{\omega^2}{\bf{\Xi}}_s^{-H}\otimes{\bf{\Xi}}_s^{-T}\otimes{\bf{A}}^{{\dagger}H}\otimes{\bf{A}}^{{\dagger}T}
\end{equation}
With (\ref{eqn:highsnrJ}), (\ref{eqn:Fisherinformmatrix}) is
rewritten into
\begin{eqnarray}
{\bf{J}}({\bf{\Xi}}_s)&=&{N_t}[{\bf{I}}_{(n_Rn_T)^2}\otimes{\text{vec}}({\bf{A}})^T][({\bf{\Xi}}_s^{-H}\otimes{\bf{\Xi}}_s^{-T})\nonumber\\
{}&{}&\otimes({\bf{A}}^{{\dagger}H}\otimes{\bf{A}}^{{\dagger}T})][{\bf{I}}_{(n_Rn_T)^2}\otimes{\text{vec}}({\bf{A}})^*]\nonumber\\
\label{eqn:Jdefnew}
{}&=&{\alpha}{N_t}({\bf{\Xi}}_s^{-H}\otimes{\bf{\Xi}}_s^{-T})
\end{eqnarray}
where
\begin{equation}
\label{eqn:alphadef}
{\alpha}={\text{vec}}({\bf{A}})^T({\bf{A}}^{{\dagger}H}\otimes{\bf{A}}^{{\dagger}T}){\text{vec}}({\bf{A}})^*=L
\end{equation}

From (\ref{eqn:Jdefnew}) and (\ref{eqn:alphadef}),
(\ref{eqn:CRLBdef}) is rewritten into
\begin{equation}
\label{eqn:CRLBdefnew}
{\text{CRLB}}({\bf{\Xi}}_s)=\frac{1}{{L}{N_t}}({\bf{\Xi}}_s^{H}\otimes{\bf{\Xi}}_s^{T})
\end{equation}
Based on (\ref{eqn:CRLBdefnew}), a lower bound of TMSE of estimating
${\bf{\Xi}}_s$ is obtained, i.e.,
\begin{equation}
\label{eqn:TMSELBdef}
{\text{TMSE}}_{LB}({\bf{\Xi}}_s)={\text{tr}}[{\text{CRLB}}({\bf{\Xi}}_s)]=\frac{(n_Tn_R)^2}{{L}{N_t}}
\end{equation}
And, accordingly, the lower bound of AvgMSE of ${\bf{\Xi}}_s$ is
\begin{equation}
\label{eqn:AvgMSELBdef}
{\text{AvgMSE}}_{LB}({\bf{\Xi}}_s)=\frac{{\text{TMSE}}_{LB}({\bf{\Xi}}_s)}{(n_Tn_R)^2}=\frac{1}{{L}{N_t}}
\end{equation}

In real applications, the number of significant eigenvalues of
${\bf{A}}$, denoted as $L_{s}$, may be less than $L$. Since the
amount of samples, $N_t$, is finite, the insignificant eigenvalues
are much less credible than significant ones. Therefore, only
significant ones are used in (\ref{eqn:MLEsecond}). Besides, the
weights of significant ones are considered to be equal. Hence,
(\ref{eqn:MLEsecond}) is changed into
\begin{eqnarray}
{}&{}&\!\!\!\!\!\!\!\!{\text{MLE}}([{\bf{\Xi}}_s]_{k_1,k_2})=\nonumber\\
\label{eqn:MLEdefnew}
{}&{}&\!\!\!\sum\limits_{l=1}^{L_{s}}\frac{[{\bf{U}}_{k_1,k_2}^H\{{\bf{\Sigma}}\}_{k_1,k_2}^{(P\times{P})}{\bf{V}}_{k_1,k_2}]_{l,l}-{\sigma_n^2}{\delta}(k_1-k_2)}{L_{s}[{\bf{\Lambda}}_{k_1,k_2}]_{l,l}}
\end{eqnarray}
Correspondingly, the lower bound of average MSE of ${\bf{\Xi}}$ is
modified into
\begin{equation}
\label{eqn:AvgMSELBdefnew}
{\text{AvgMSE}}_{LB}({\bf{\Xi}}_s)=\frac{1}{{\alpha}{N_t}}=\frac{1}{{L_s}{N_t}}
\end{equation}
In fact, $L_s$ represents the order of frequency selectivity, that
is, the number of equivalent independent parallel transmission
branches of the multipath channels for OFDM systems.

When the SNR is finite, ${\bf{\Sigma}}'$ is approximated as
\begin{equation}
\label{eqn:Sigmaprimefinitesnr}
{\bf{\Sigma}}'=\frac{\omega}{N_t}{\bf{\Xi}}_s\otimes({\bf{A}}+\frac{\sigma_n^2}{\omega}{\bf{I}}_{P})
\end{equation}
and, correspondingly, (\ref{eqn:CRLBdef}) is  rewritten into
\begin{equation}
\label{eqn:CRLBdefnew2}
{\text{CRLB}}({\bf{\Xi}}_s)=\frac{1}{{\beta}{N_t}}({\bf{\Xi}}_s^{H}\otimes{\bf{\Xi}}_s^{T})
\end{equation}
and (\ref{eqn:AvgMSELBdef}) is rewritten into
\begin{equation}
\label{eqn:AvgMSELBdefnew2}
{\text{AvgMSE}}_{LB}({\bf{\Xi}}_s)=\frac{1}{{\beta}{N_t}}
\end{equation}
where
\begin{eqnarray}
{\beta}&=&{\text{vec}}({\bf{A}})^T[({\bf{A}}+\frac{\sigma_n^2}{\omega}{\bf{I}}_{P})^{-H}\otimes({\bf{A}}+\frac{\sigma_n^2}{\omega}{\bf{I}}_{P})^{-T}]{\text{vec}}({\bf{A}})^*\nonumber\\
\label{eqn:betadef}
{}&=&\sum\limits_{l=1}^{L}[\frac{1}{1+(\omega\rho_l)^{-1}}]^2
\end{eqnarray}
where $\rho_l=\frac{\lambda_l}{\sigma_n^2}$ and $\lambda_l$ is the
$l$-th eigenvalue of ${\bf{A}}$. Moreover, since
\begin{equation}
\label{eqn:omegafirst}
\omega={\bf{x}}_p^H{\bf{\Omega}}{\bf{x}}_p=\|{\bf{x}}_p\|_2^2\times\frac{{\bf{x}}_p^H{\bf{\Omega}}{\bf{x}}_p}{{\bf{x}}_p^H{\bf{x}}_p}=\|{\bf{x}}_p\|_2^2\times{\text{R}}({\bf{\Omega}})
\end{equation}
where $\|{\bf{x}}_p\|_2^2$ is the power of pilot symbol, and
${\text{R}}({\bf{\Omega}})$ is the Rayleigh quotient of
${\bf{\Omega}}$ \cite{Golu96}. Due to the normalized power,
$\|{\bf{x}}_p\|_2^2=P$. Besides, it is straightforward that
${\text{R}}({\bf{\Omega}})\le{\lambda}_{max}({\bf{\Omega}})$, where
${\lambda}_{max}({\bf{\Omega}})$ denotes the maximum eigenvalue of
${\bf{\Omega}}$. According to \cite{ZhaoICC09CRLB}, when
${\theta}f_dT_s\le0.35$, ${\lambda}_{max}({\bf{\Omega}})$ can be
well approximated by
\begin{equation}
{\lambda}_{max}({\bf{\Omega}})\;{\approx}\;PJ_0(2{\pi}c{\theta}f_dT_s)
\end{equation}
where $c=0.35$. Therefore, $\beta$ is upper bounded by
\begin{equation}
\label{eqn:betadefnew}
\beta\leq\sum\limits_{l=1}^{L}[\frac{1}{1+(P^2J_0(2{\pi}c{\theta}f_dT_s)\rho_l)^{-1}}]^2=\beta_{max}
\end{equation}
Then, (\ref{eqn:AvgMSELBdefnew2}) is further lower bounded by
\begin{equation}
\label{eqn:AvgMSELBdefnew3}
{\text{AvgMSE}}_{LB}({\bf{\Xi}}_s)=\frac{1}{{\beta_{max}}{N_t}}
\end{equation}
Note that $\omega\rho_l$ can be regarded as the effective SNR on the
$l$-th subchannel. When $\omega\rho_l$ is too small, say, below 0
dB, it should not be used in MLE (\ref{eqn:MLEdefnew}), which would
reduce the effective order of frequency selectivity. According to
(\ref{eqn:betadefnew}), therefore, the number of pilot tones, SNR
and maximum Doppler spread together influence the effective order of
frequency selectivity and, further, the accuracy of estimation.

\begin{figure}
\centering{\subfloat[EVA]{\includegraphics[width=3.2in,height=1.9in]{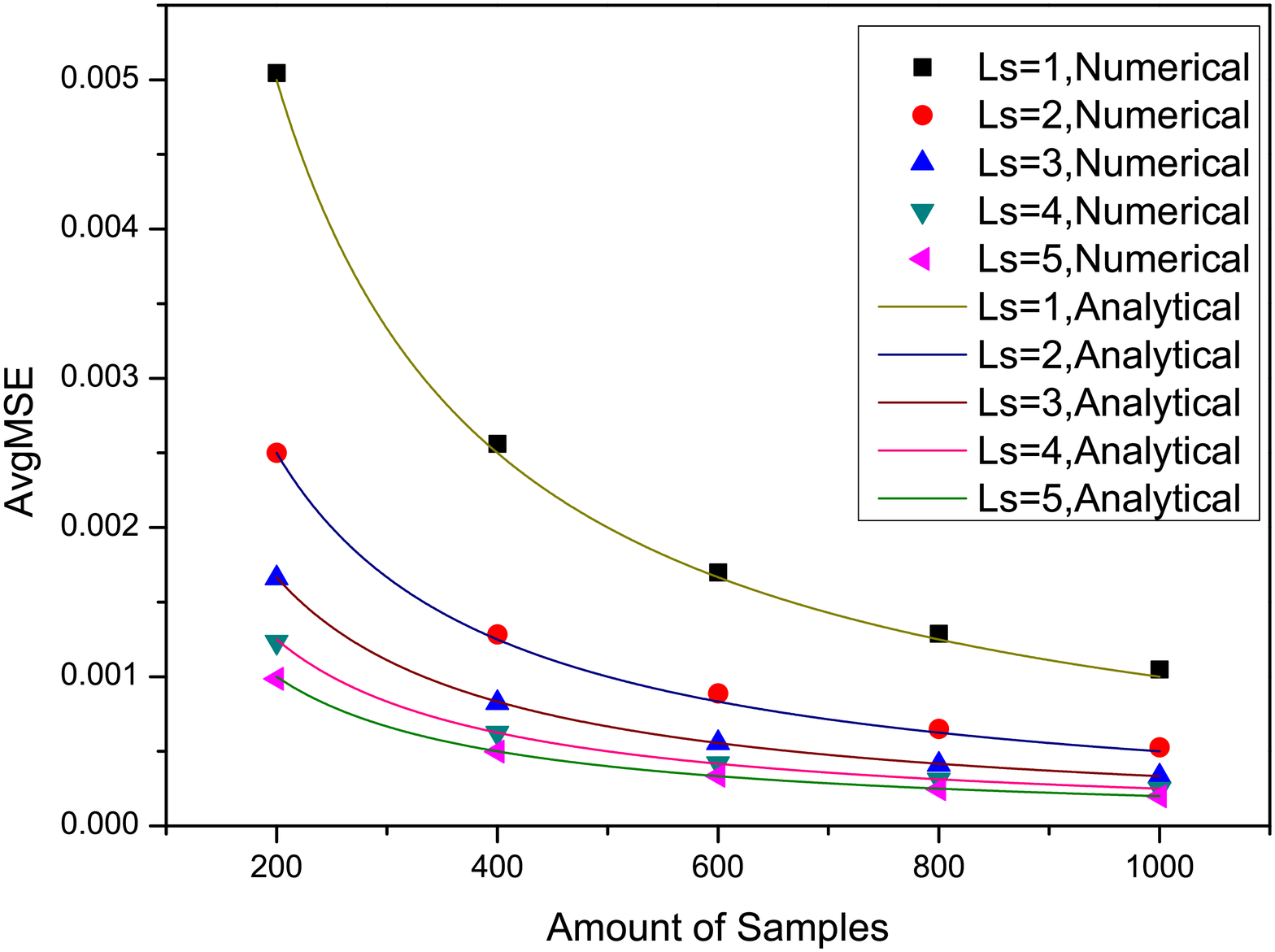}
\label{subfig:evasnr}} \\
\subfloat[ETU]{\includegraphics[width=3.2in,height=1.9in]{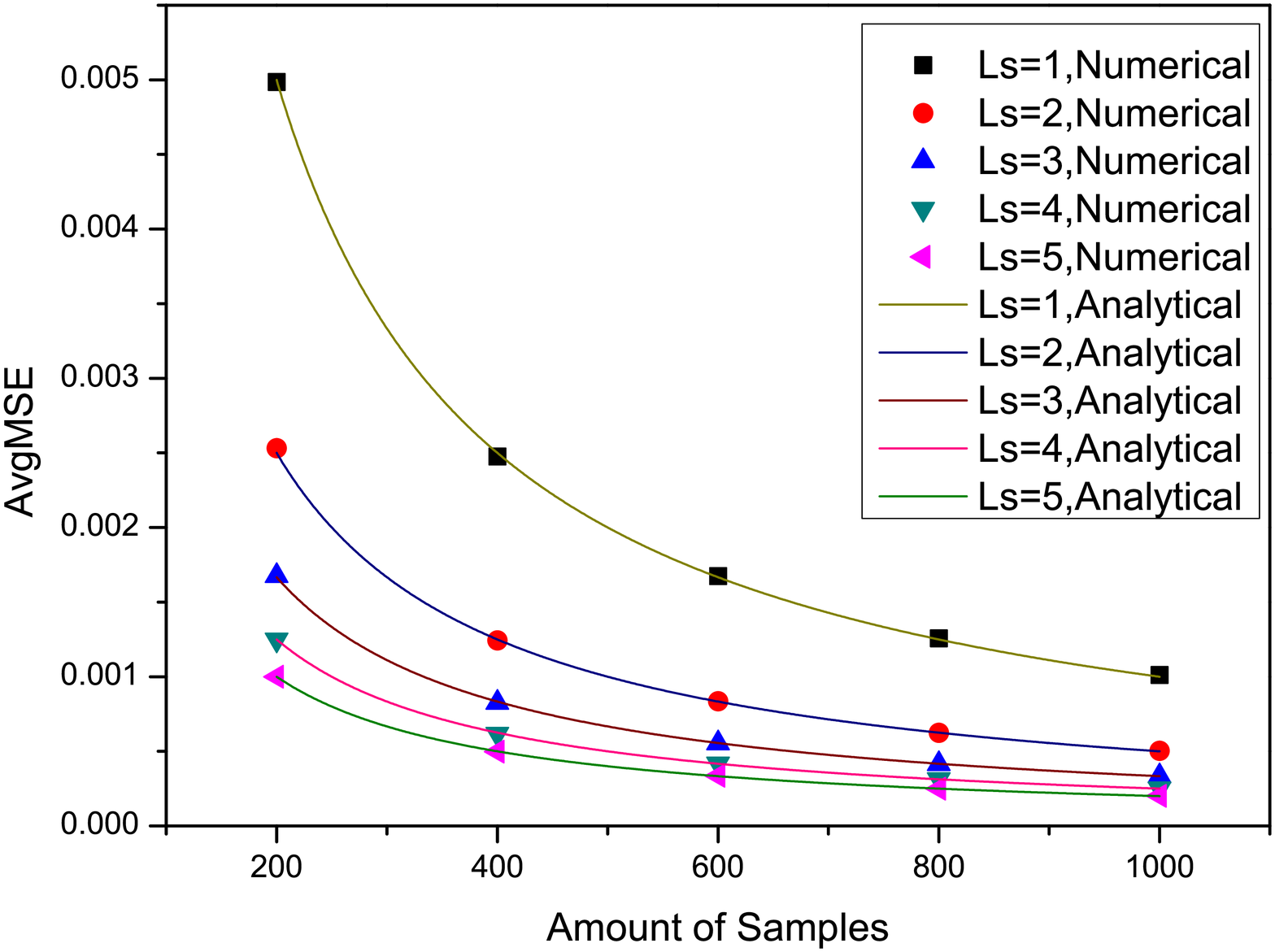}
\label{subfig:etusnr}}} \caption{Comparison of analytic results of
(\ref{eqn:AvgMSELBdefnew}) and numerical results for EVA and ETU
channels when $SNR\rightarrow\infty$ and $f_d=100$Hz.}
\label{fig:fig1}
\end{figure}

\section{Numerical Results}
\label{sec:numresults} The OFDM system in simulations is of
$BW=1.25$ MHz ($T=1/BW=800$ ns), $N=128$, and $L_{cp}=16$. Two 3GPP
E-UTRA channel models are adopted: Extended Vehicular A model (EVA)
and Extended Typical Urban model (ETU) \cite{3GPP36101}. The excess
tap delay of EVA is [$0$, $30$, $150$, $310$, $370$, $710$, $1090$,
$1730$, $2510$] ns, and its relative power is [$0.0$, $-1.5$,
$-1.4$, $-3.6$, $-0.6$, $-9.1$, $-7.0$, $-12.0$, $-16.9$] dB. For
ETU, they are [$0$, $50$, $120$, $200$, $230$, $500$, $1600$,
$2300$, $5000$] ns and [$-1.0$, $-1.0$, $-1.0$, $0.0$, $0.0$, $0.0$,
$-3.0$, $-5.0$, $-7.0$] dB, respectively. The classic Doppler
spectrum, i.e., Jakes' spectrum \cite{Stee92}, is applied to
generate the Rayleigh fading channel. The MIMO configuration is
$4\times4$, and the correlation matrices of transmitting and
receiving antennas are shown at the bottom of the next page,
respectively \cite{Kerm02}. Besides, the number of pilot tones per
transmitting antenna is 16, the pilot spacing is 8, and the pilot
symbols are separated far enough from others to reduce the
correlation among them.

\begin{figure*}[!t]
\centering{\subfloat[EVA,$f_d=40$Hz.]{\includegraphics[width=2.3in,height=1.5in]{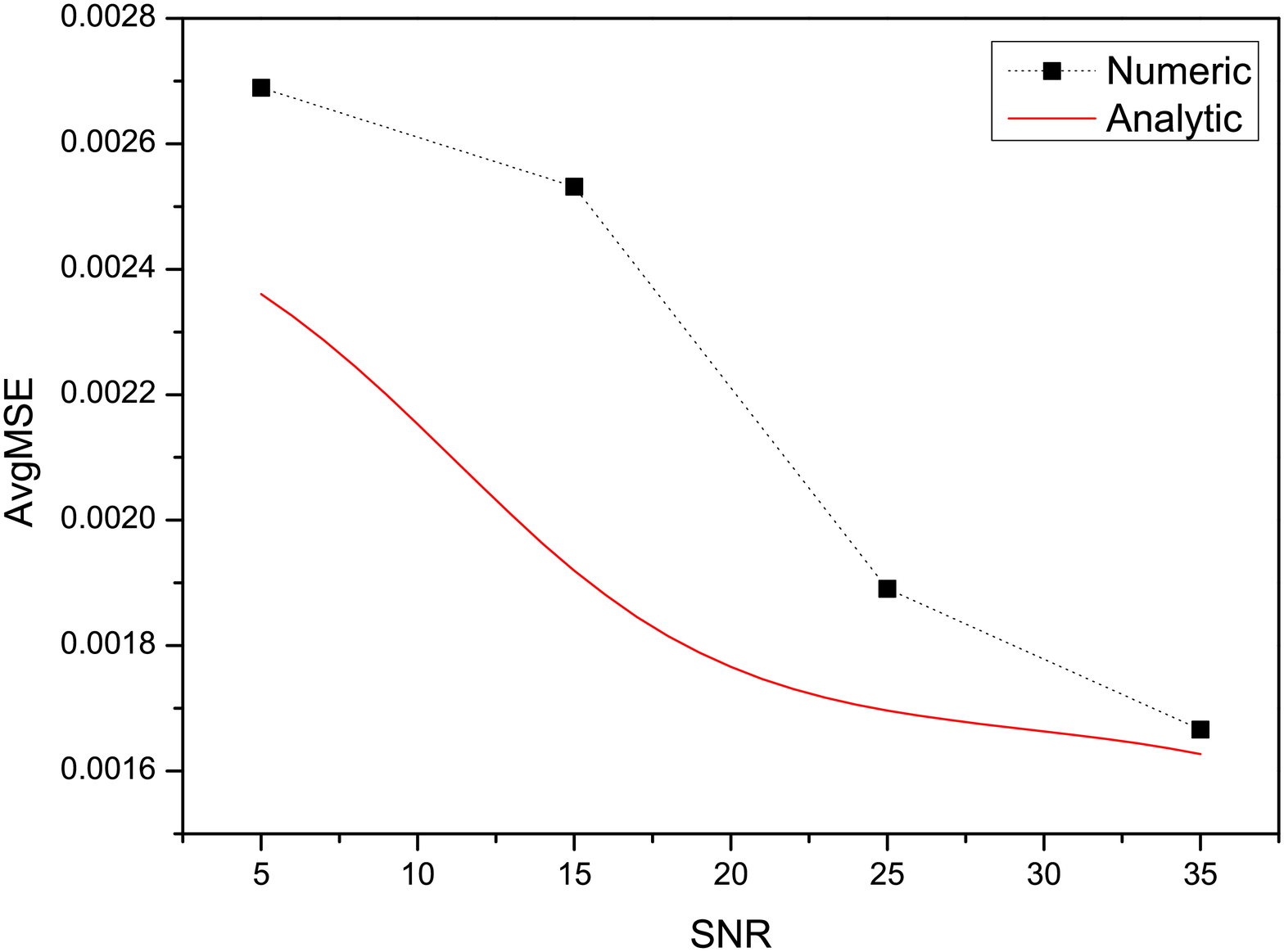}
\label{subfig:evafd40}} \hfil
\subfloat[EVA,$f_d=80$Hz.]{\includegraphics[width=2.3in,height=1.5in]{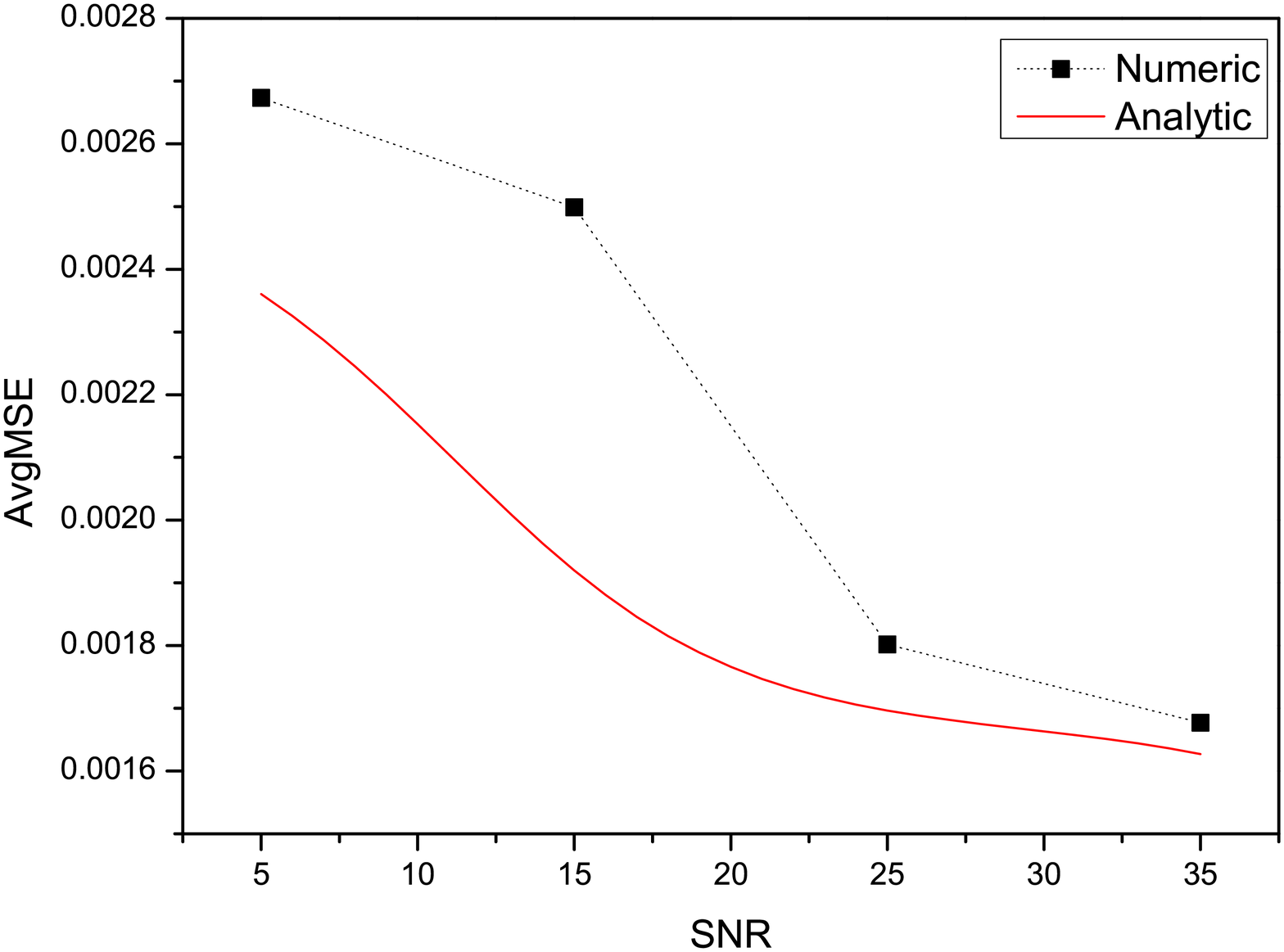}
\label{subfig:evafd80}} \hfil
\subfloat[EVA,$f_d=120$Hz.]{\includegraphics[width=2.3in,height=1.5in]{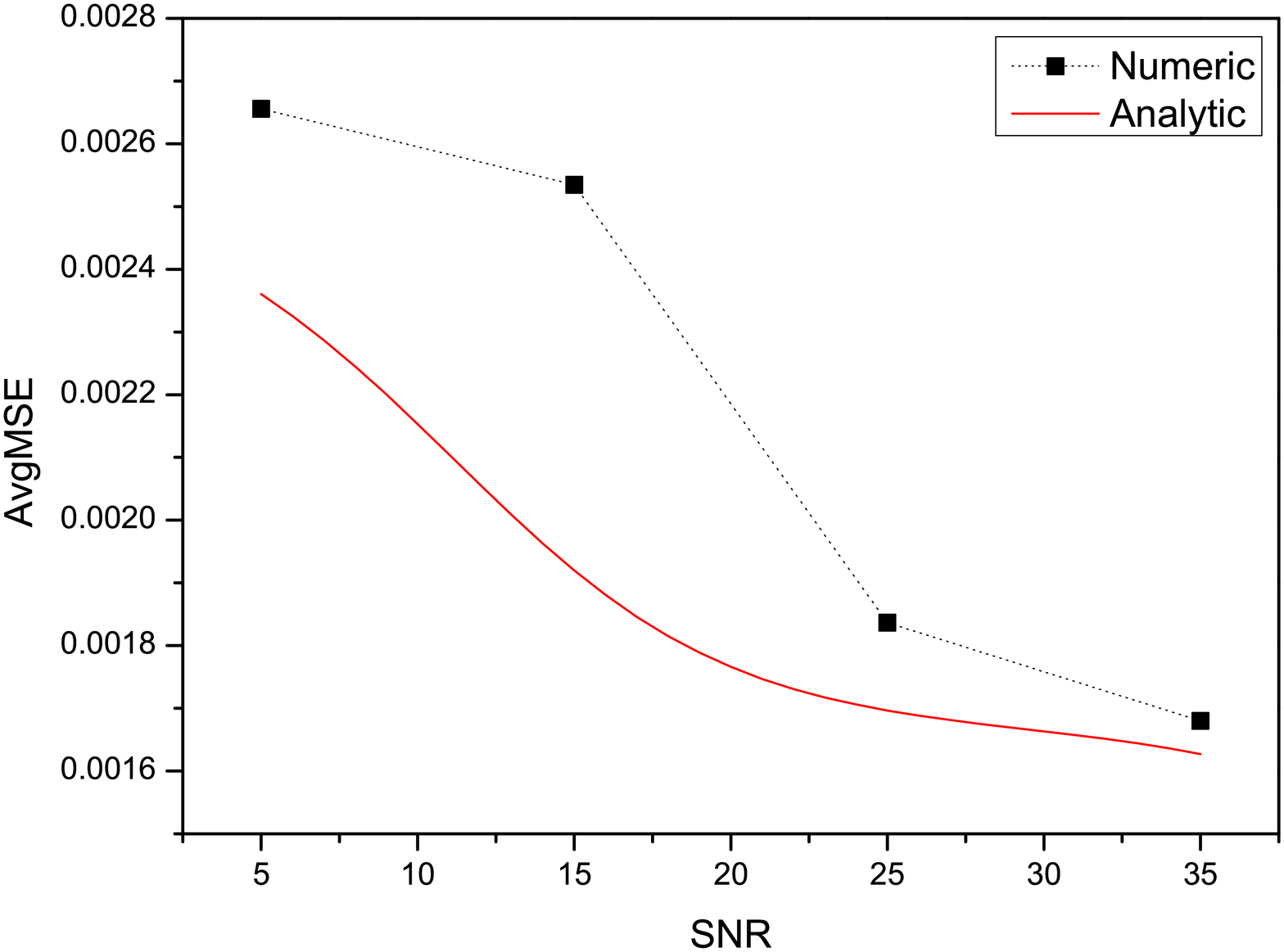}
\label{subfig:evafd120}}} \\
\centering{\subfloat[ETU,$f_d=40$Hz.]{\includegraphics[width=2.3in,height=1.5in]{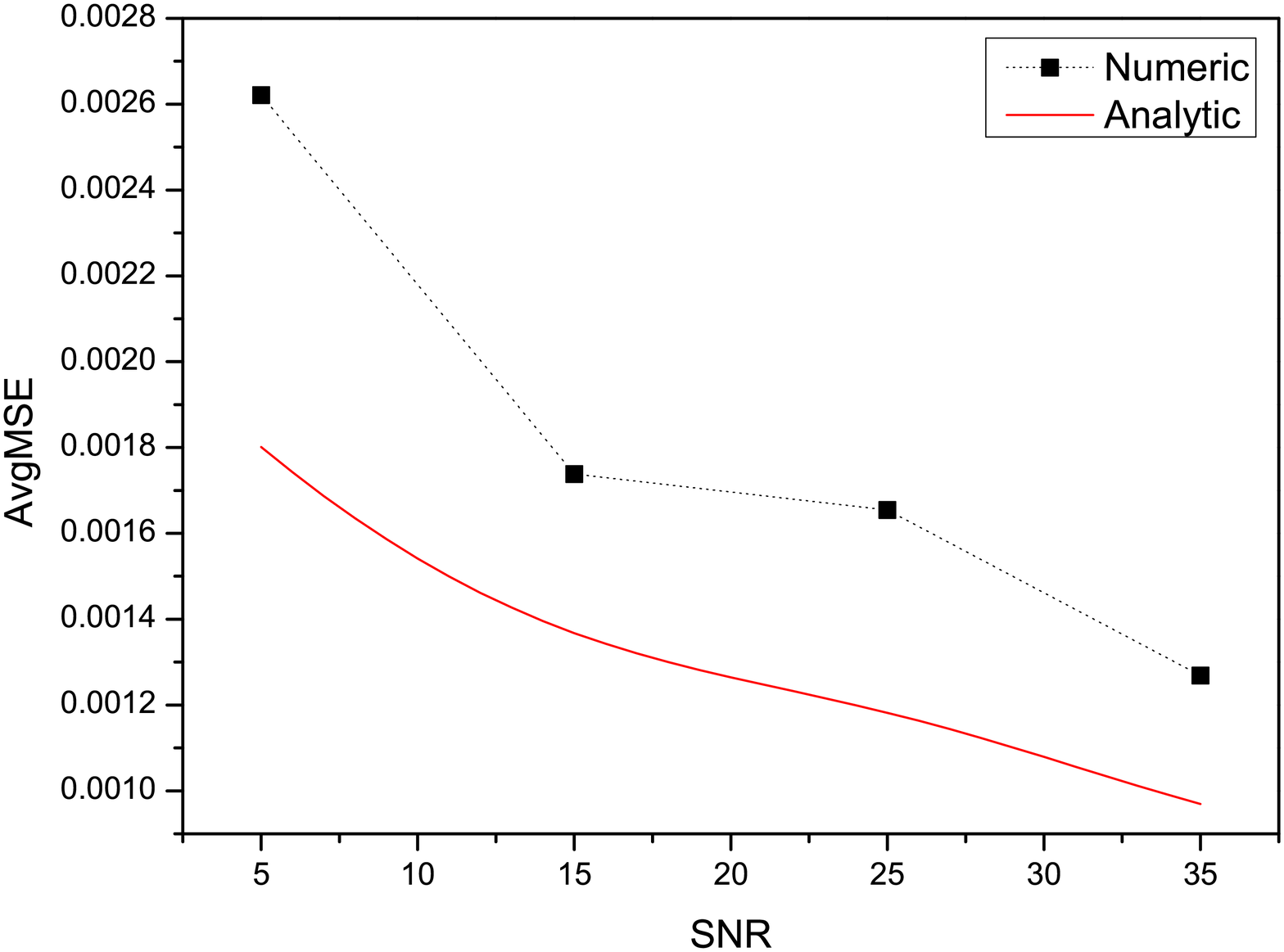}
\label{subfig:etufd40}} \hfil
\subfloat[ETU,$f_d=80$Hz.]{\includegraphics[width=2.3in,height=1.5in]{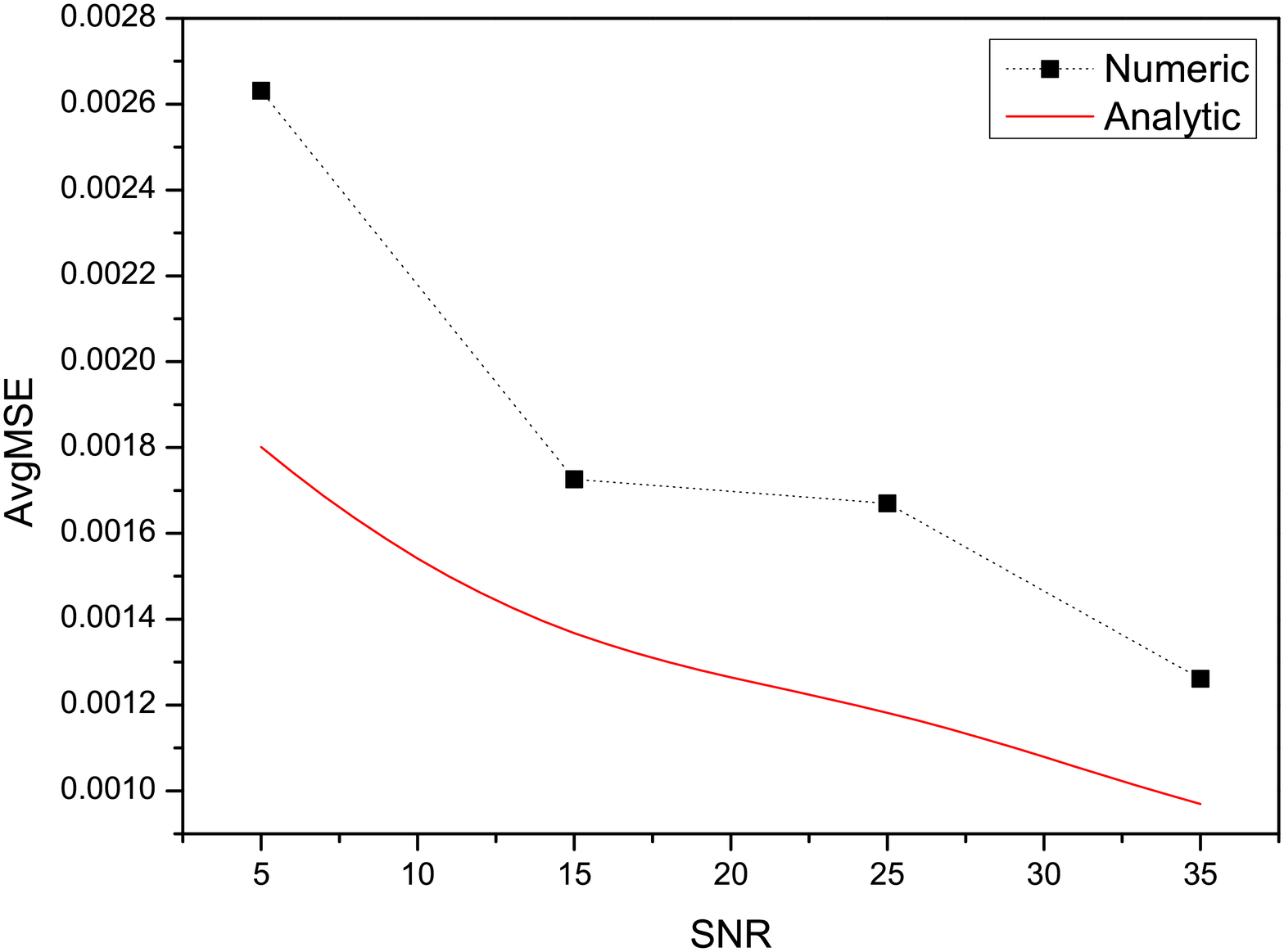}
\label{subfig:etufd80}} \hfil
\subfloat[ETU,$f_d=120$Hz.]{\includegraphics[width=2.3in,height=1.5in]{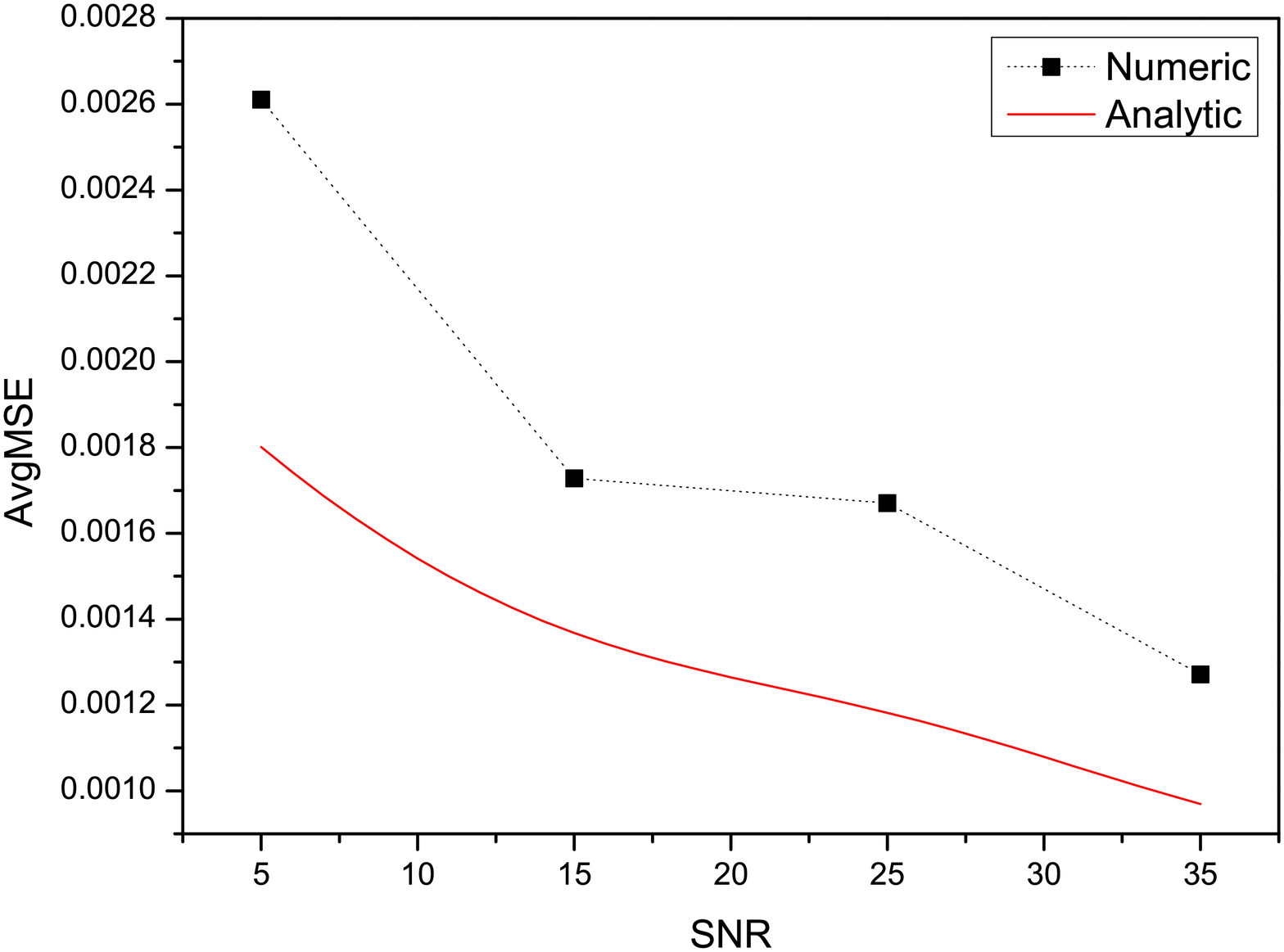}
\label{subfig:etufd120}}} \caption{Comparison of analytic results of
(\ref{eqn:AvgMSELBdefnew3}) and numerical results for EVA and ETU
channels under different SNR's and Doppler's.} \label{fig:fig2}
\end{figure*}

\begin{figure*}[!b]
\vspace{-4pt} \hrule \vspace{4pt} {\small{
\begin{array}{ll}
{\bf{\Xi}}_{s,T}= &
{\bf{\Xi}}_{s,R}= \\
\left(
                   \begin{array}{cccc}
         1      & -0.13-j0.62 & -0.49+j0.23 &  0.15+j0.28 \\
    -0.13+j0.62 &      1      & -0.13-j0.52 & -0.38+j0.12 \\
    -0.49-j0.23 & -0.13+j0.52 &      1      &  0.02-j0.61 \\
     0.15-j0.28 & -0.38-j0.12 &  0.02+j0.61 &      1
                   \end{array}
                 \right), &
\left(
                   \begin{array}{cccc}
         1      & -0.45+j0.53 &  0.37-j0.22 &  0.19+j0.21 \\
    -0.45-j0.53 &      1      & -0.35-j0.02 &  0.02-j0.27 \\
     0.37+j0.22 & -0.35+j0.02 &      1      & -0.10+j0.54 \\
     0.19-j0.21 &  0.02+j0.27 & -0.10-j0.54 &      1
                   \end{array}
                 \right)
\end{array}}}
\end{figure*}

In Fig.\ref{fig:fig1}, we compare the analytic results
(\ref{eqn:AvgMSELBdefnew}) and the numerical results over a range of
$N_t$'s for EVA and ETU channels, respectively, when the SNR is
asymptotically infinite and $f_d=100$Hz. The pilot sequences are
QPSK modulated and randomly chosen. Besides, different $L_s$'s are
tested to demonstrate that (\ref{eqn:AvgMSELBdefnew}) is a tight
lower bound of (\ref{eqn:MLEdefnew}). Apparently, the analytic
results meet the numerical ones quite well.

In Fig.\ref{fig:fig2}, different SNR's and maximum Doppler spreads
are tested to demonstrate that (\ref{eqn:AvgMSELBdefnew3}) is tight
lower bound of (\ref{eqn:MLEdefnew}) when the pilot sequence,
${\bf{x}}_p$, is properly selected, specifically, the eigenvector of
${\bf{\Omega}}$ associated with the maximum eigenvalue. Since the
number of pilot tones, SNR and normalized maximum Doppler spread
together influence the effective order of frequency selectivity, the
value of $L_s$ varies for different cases, which is reflected by
varying AvgMSE. Further, it is obvious that SNR has a more
significant impact on $L_s$ than the maximum Doppler spread, when
the number of pilot tones are large enough, e.g., over 16.

\section{Conclusion}
\label{sec:conclusion} In this paper, we derive the CRLB of spatial
correlation matrices based on a rigorous model of the doubly
selective fading channel for MIMO OFDM systems. With several
necessary assumptions, the sample auto-correlation matrix of the
channel response is complex Wishart distributed. Then, the maximum
likelihood estimator is obtained, and the analytic expressions of
CRLB as well as lower bounds of TMSE and AvgMSE are deduced for
asymptotically infinite and finite SNR's, respectively. According to
the lower bound of AvgMSE, the amount of samples and the order of
frequency selectivity influence the accuracy of estimation
dominantly. Besides, the number of pilot tones, SNR and maximum
Doppler spread have effects on the effective order of frequency
selectivity.

\setlength{\arraycolsep}{5pt}

\bibliographystyle{IEEEtran}
\bibliography{IEEEabrv,myBibs}

\end{document}